\documentstyle[11pt,aaspp4,psfig]{article}
\def\pp{\par\parshape 2 0truecm 15.5truecm 1truecm 14.5truecm\noindent}
\newcommand{\simgt}{\lower.5ex\hbox{$\; \buildrel > \over \sim \;$}}
\newcommand{\simlt}{\lower.5ex\hbox{$\; \buildrel < \over \sim \;$}}
\newcommand{\himpc}{{\hbox {$h^{-1}$}{\rm Mpc}} }
\begin{document}
\begin{minipage}[c]{4cm}
RESCEU-47/97\\
UTAP-278/97\\
ADAC-006/1997\\
\end{minipage}\\

\title{Hydrodynamical Simulation of Clusters of Galaxies \\
in X--Ray, mm, and submm Bands:\\
Determination of Peculiar Velocity and the Hubble Constant}

\bigskip

\author{Kohji Yoshikawa$^{1}$, Makoto Itoh$^{2}$ and Yasushi Suto$^{3,4}$}

\bigskip
\bigskip

\affil{
$^{1}$ Department of Astronomy, Kyoto University, Kyoto 606-8502, Japan.\\
$^{2}$ Center for Information and Multimedia Studies, 
Kyoto University, Kyoto 606-8501, Japan.\\
$^{3}$ Department of Physics, The University of Tokyo, 
Tokyo 113-0033, Japan.\\
$^{4}$ Research Center For the Early Universe (RESCEU), 
School of Science, \\
The University of Tokyo, Tokyo 113, Japan.\\
~\\
e-mail: kohji@kusastro.kyoto-u.ac.jp, 
mitoh@media.kyoto-u.ac.jp, 
suto@phys.s.u-tokyo.ac.jp}

\received{} \accepted{}

\begin{abstract}
  We have performed a series of simulations of clusters of galaxies on
  the basis of the smoothed particle hydrodynamics technique in a
  spatially-flat cold dark matter universe with $\Omega=0.3$,
  $\lambda=0.7$, and $H_0=70$km/s/Mpc as one of the most successful
  representative cosmological scenarios. In particular, we focus on
  the Sunyaev--Zel'dovich effect in submm and mm bands, and estimate
  the reliability of the estimates of the global Hubble constant $H_0$
  and the peculiar velocity of clusters $v_r$. Our simulations
  indicate that fractional uncertainties of the estimates of $H_0$
  amount to $\sim 20$\% mainly due to the departure from the
  isothermal and spherical gas density distribution. We find a
  systematic underestimate bias of $H_0$ by $\sim 20$\% for clusters
  $z\approx 1$, but not at $z\approx 0$. The gas temperature drop in
  the central regions of our simulated clusters leads to the
  underestimate bias of $v_r$ by $\sim 5$\% at $z\approx 0$ and by
  $\sim 15$\% at $z\approx 1$ in addition to the statistical errors of
  the comparable amount due to the non-spherical gas profile.
\end{abstract}


\vfill

\centerline{\sl Publications of the Astronomical Society of Japan
(1998), in press}

\newpage


\section{Introduction}

Clusters of galaxies have been extensively observed in radio, optical
and X--ray bands. Furthermore recent and future observational
facilities in mm and submm bands, such as the SCUBA (Submillimeter
Common-User Bolometer Array), the Japanese LMSA (Large Millimeter and
Submillimeter Array) project and the European PLANCK mission are
expected to open the submm window to observe clusters of galaxies via
the Sunyaev--Zel'dovich (SZ) effect (Sunyaev, Zel'dovich 1972) in
addition to the Rayleigh-Jeans region of the spectrum of the cosmic
microwave background (CMB) where the SZ temperature decrement is
reported for about a dozen of clusters (e.g., Rephaeli 1995;
Kobayashi, Sasaki, Suto 1996). Since the intensity of the SZ effect
does not suffer from the $(1+z)^{-4}$ diminishing factor unlike X--ray
surface brightness, observations in mm and submm bands are much more
advantageous for clusters, especially at high $z$, than those in
optical and X-ray bands (Barbosa et al. 1996; Silverberg et al. 1997;
Kitayama, Sasaki, Suto 1998).

As extensively discussed in previous literatures (Silk, White 1978;
Sunyaev, Zel'dovich 1980; Birkinshaw, Hughes, Arnaud 1991; Rephaeli,
Lahav 1991; Inagaki, Suginohara, Suto 1995; Kobayashi, Sasaki, Suto
1996; Holzapfel et al. 1997; Kitayama, Sasaki, Suto 1998), one can
then combine these multi-band observations of high--$z$ clusters to
determine the cosmological parameters and the peculiar velocity of
clusters.  These procedures, however, usually assume that the gas of
clusters is isothermal and spherical, while all the observed clusters
do exhibit a departure from the assumption to some extent. The
departure would hamper the reliable estimates of, for instance, the
Hubble constant $H_0$ and the peculiar velocity $v_r$.  To address the
question quantitatively, we carry out a series of numerical
simulations of clusters. We extract simulated clusters both at
$z\approx 0.0$ and $z\approx 1.0$, and perform the ``simulated''
observations in X--ray, mm and submm bands. Finally we combine the
multi-band information to evaluate the statistical and possible
systematic errors of the estimates of $H_0$ and $v_r$. Our present
work extends the previous studies of this methodology (Inagaki,
Suginohara, Suto 1995; Roettiger, Stone, Mushotzky 1997) for $H_0$ and
examines uncertainties of the peculiar velocity field as well, paying
attention to the projection effect and the evolution of clusters.


\section{Numerical Simulation}

As a representative cosmological model, we consider a cold dark matter
cosmogony with $\Omega_0=0.3$, $\lambda_0=0.7$, $h=0.7$,
$\sigma_8=1.0$, $n=1$ and $\Omega_b=0.015h^{-2}$, where $\Omega_0$ is
the density parameter, $\lambda_0$ is the cosmological constant, $h$
is $H_0$ in units of 100km/s/Mpc, $\sigma_8$ is the amplitude of the
density fluctuation, $n$ is the spectral index of the primordial
fluctuations, and $\Omega_b$ is the baryon density parameter. This
model satisfies both the COBE normalization and the cluster abundance
(White, Efstathiou, Frenk 1993; Kitayama, Suto 1997), and also is
consistent with the recent indications of no significant evolution of
galaxy clusters (see Kitayama, Sasaki, Suto 1998 for details).

The initial conditions of our simulations are generated using the
COSMICS package by E. Bertschinger and the resulting gas and dark
matter particle distributions are evolved with the publicly available
AP$^3$M--SPH code ({\it Hydra}) by Couchman et al. (1995).  The
effects of radiative cooling and heating are neglected. We use the
ideal gas equation of state with an adiabatic index $\gamma=5/3$.
Gravitational forces are softened with a physical softening length of
$\epsilon_{\rm G} = 39h^{-1}$kpc.

Our simulations proceed in two steps; first we carry out three
low--resolution simulations with $N=64^3$ particles each for gas and
dark matter in a comoving periodic cubic of $L_{\rm box}=$ 100\himpc
(one realization) and 200\himpc(two realizations). Then we identify
clusters at $z=0$ using the friend--of--friend algorithm with a
bonding length of 0.2 times the mean particle separation.  The virial
mass of clusters is computed from all the particles within the virial
radius, $r_{\rm vir}$, from the center of each cluster. The latter is
defined so that the mean density inside becomes $\rho_{\rm vir}\simeq
18\pi^2 \Omega_0^{0.4} \overline{\rho}_c(z)$, the virialized density
at $z$ predicted in the spherical non--linear model (e.g., White,
Efstathiou, Frenk 1993; Kitayama, Suto 1996), where
$\overline{\rho}_c(z)$ is the critical density of the universe at
redshift of $z$. The resulting mass function for the identified
clusters from the low--resolution simulations is consistent, within a
factor of two, with the theoretical Press--Schechter mass function
(Press, Schechter 1974).  Considering a somewhat simplified
identification scheme adopted here, the agreement is satisfactory and
implies that our low--resolution simulations provide fairly
homogeneous and unbiased catalogues of clusters.

From the cluster catalogues constructed by the three low--resolution
simulations, we select nine clusters with mass greater than $10^{14}
M_{\odot}$. Then we set a box of side 25\himpc (5 clusters), 50\himpc
(3 clusters) or 100\himpc (1 cluster) at the center of each cluster
and fill the box with $N=64^3$ particles each for gas and dark matter
which are assigned the same amplitude and phases of perturbation waves
as in the initial condition of the low--resolution simulation at
$z=32$. The nine initial conditions for high--resolution simulations
are evolved again with the periodic boundary condition.


\section{Fluxes in the X-ray, mm and submm bands}

The observable quantities of clusters which we consider in this letter
are the X--ray surface brightness and the spectral intensities due to
the thermal and kinematic SZ effect at mm and submm bands. The X--ray
surface brightness of clusters between the frequency bands $\nu_1$ and
$\nu_2$ for clusters located at $z$ is given by the following integral
along the line--of--sight (e.g., Rybicki, Lightman 1979):
\begin{equation}
  \label{eqn:x-ray}
  S_{\rm X}(\nu_1,\nu_2) = \frac{1}{4\pi(1+z)^4}\int^{\infty}_{-\infty}
  \alpha[\nu_1(1+z),\nu_2(1+z), T_e]\,n_e^2\,dl,
\end{equation}
where $\alpha[\nu_1,\nu_2,T_e]$ is the X--ray emissivity in the
corresponding frequency band, $T_e$ is the temperature of electron gas
and $n_e$ is the number density of electrons. In practice we adopt the
Masai model (Masai 1984) for $\alpha[\nu_1,\nu_2,T_e]$ and thus
include both metal line emissions and the thermal bremsstrahlung; the
former becomes important for low temperature clusters.

The spectral intensity due to the thermal SZ effect is written as
(e.g., Rephaeli 1995):
\begin{equation}
  \label{eqn:dIth}
  \Delta I^{\rm th}_{\nu}=i_0 y\: g(x),
\end{equation}
where $i_0\equiv2(k_{\rm B}T_{\rm CMB})^3/(h_pc)^2$, $y$ is the
Compton $y$--parameter :
\begin{equation}
  \label{eqn:yint}
  y=\int_{-\infty}^{\infty}\frac{k_{\rm B}T_{\rm e}}{m_{\rm e}c^2}
  \sigma_{\rm T}n_{\rm e}\,dl,
\end{equation}
and $g(x)$ is a function of $x \equiv h_p\nu/k_{\rm B}T_{\rm CMB}$: 
\begin{equation}
  \label{eqn:gx}
  g(x) = \frac{x^4 e^x}{(e^x-1)^2}
  \left[x\coth\left(\frac{x}{2}\right)-4\right].
\end{equation}
In the above expressions, $k_{\rm B}$, $m_{\rm e}$ $\sigma_{\rm T}$,
$c$, $h_p$ and $T_{\rm CMB}$ denote the Boltzmann constant, the
electron mass, the Thomson cross section, the velocity of light, the
Planck constant, and the temperature of the CMB, respectively.

The corresponding change of the CMB temperature due to the
thermal SZ effect is
\begin{equation}
  \label{eqn:dToverT}
  \frac{\Delta T^{\rm th}}{T_{\rm CMB}}=y\,t(x),
\end{equation}
where $t(x)$ is given by
\begin{equation}
  \label{eqn:tx}
  t(x)=\frac{x^2e^x}{(e^x-1)^2}
\left[x\coth\left(\frac{x}{2}\right)-4\right].
\end{equation}

The peculiar velocity of a cluster along the line--of--sight, $v_r$,
relative to the CMB rest frame (defined to be positive for a receding
cluster) produces a kinematic SZ effect (Sunyaev, Zel'dovich 1980):
\begin{equation}
  \label{eqn:dIkin}
  \Delta I^{\rm kin}_{\nu}=-i_0\frac{v_r}{c}\tau h(x)
  =-i_0\frac{v_r}{c}\frac{m_{\rm e}c^2}{k_{\rm B}T_{\rm e}}y \,h(x),
\end{equation}
where $\tau$ is an optical depth of the cluster, and
\begin{equation}
  \label{8}
  h(x)\equiv\frac{x^4e^x}{(e^x-1)^2}.
\end{equation}
Note that the last equality of equation (\ref{eqn:dIkin}) assumes that
the temperature profile of the cluster is isothermal, the validity of
which we will discuss further in \S 4.2. 

\begin{table}[tbph]
  \begin{center}
    \leavevmode
    \begin{tabular}{l|c|c|c|c|}
      &\multicolumn{2}{c|}{cluster A}&\multicolumn{2}{c|}{cluster B}
 \\ \cline{2-5}
cluster quantities &$z=0$ &$z=1$ &$z=0$ &$z=1$  \\ \hline \hline 
$r_{\rm vir}$\,[Mpc] & 2.0 & 0.80 & 0.80 & 0.40\\ 
$M_{\rm vir} \, [M_{\odot}]$
      &$1.3\times 10^{15}$ & $5.5\times 10^{14}$ 
      &$1.2\times10^{14}$ & $6.1\times 10^{13}$ \\ 
$N_{\rm gas}(<r_{\rm vir})$ & 2186 & 819 & 14501 & 6957\\ 
$N_{\rm DM}(<r_{\rm vir})$ & 2801 & 1191 & 17379 & 8482\\ 
$L_{\rm X,2-10}$ \,[10$^{44}$erg/sec] & 11 & 10 & 0.85 & 1.2 \\ 
$\overline{T}_{\rm X}$\,[keV] &7.1 & 5.9 & 2.1 & 1.4\\ 
$\sigma_{\rm 1D}$\,[km/sec]  & 1300 & 1100 & 587 & 545 \\ 
$\beta_{\rm spec}$ &1.4& 1.3& 1.0 & 1.3\\ \cline{1-5}
$L_{\rm box}$\, [$\himpc$]&\multicolumn{2}{c|}{$100$}&\multicolumn{2}{c|}{$25$}
    \end{tabular}
\caption{Physical quantities characterizing the simulated clusters A and B 
at $z=0.0$ and $z=1.0$: the virial radius in physical lengths ($r_{\rm
vir}$), the total mass within the virial radius ($M_{\rm vir}$), the
number of gas and dark matter particle within the virial radius
($N_{\rm gas}$ and $N_{\rm DM}$), the X--ray luminosity in 2--10 keV
band ($L_{\rm X,2-10}$), the X-ray emission weighted temperature
($\overline{T}_{\rm X}$), the 1--dimensional velocity dispersion of
dark matter ($\sigma_{\rm 1D}$), $\beta$--parameter defined by
$\beta_{\rm spec}=\mu m_{\rm p}\sigma^2_{\rm 1D}/k_{\rm B} \overline
{T}_{\rm X}$, and $L_{\rm box}$ is the comoving size of the simulation
box.}
    \label{tab:1}
  \end{center}
\end{table}

From the nine simulated clusters in total, we trace their progenitors
at $z=1.0$.  We choose two clusters, A and B, so as to represent rich
and poor clusters in our simulated sample, respectively.
Table~\ref{tab:1} summarizes their properties at $z=0$ and $z=1.0$, in
which the X--ray emission weighted temperature is defined as
\begin{equation}
  \label{eqn:xemte}
  \overline{T}_X  \equiv 
  \frac{\displaystyle \int_0^{r_{\rm vir}} T_{\rm e}(r) 
              \alpha[T_e] n_{e}^2(r) r^2dr }
  {\displaystyle \int_0^{r_{\rm vir}} \alpha[T_e] n_{e}^2(r) r^2dr}.
\end{equation}
Figure \ref{fig:2dprofile} shows the projected profiles of the X-ray
surface brightness, temperature decrement, and submm surface
brightness at $z=0.05$ and $z=1.0$ for the two clusters, and
dash-dotted lines show the $\beta$--model profiles adopting their
fitted parameters. The energy bands adopted in Figure
\ref{fig:2dprofile} correspond to those of ASCA (2-10keV) in X--ray,
NOBA (NOBeyama Bolometer Array; 150GHz) in mm, and SCUBA (350GHz) in
submm. Note that we are mainly interested in quantifying the {\it
fractional} uncertainties in the measurements of $H_0$ and the
peculiar velocities. In this sense, the absolute magnitude of the
value of the $y$-parameter is not important; since our simulations did
not include radiative cooling and other scale-dependent physical
processes, the results are almost scalable and it seems unlikely that
our results presented here change drastically for much larger clusters
which have the $y$-parameter large enough to be observable in
practice.  We plan to come back to the issue with much larger
simulations including various scale-dependent physical processes in
due course.

If the cluster gas strictly obeys the isothermal $\beta$--model:
\begin{equation}
  \label{eqn:betamodel}
  n_{\rm e}(r) = n_{\rm e0}[1+(r/r_{\rm c})^2]^{-3\beta/2},
\end{equation}
then X--ray surface brightness (eq.[\ref{eqn:x-ray}]) and y-parameter
(eq.[\ref{eqn:yint}]) should be given by
\begin{equation}
  \label{eqn:Sx}
  S_{\rm X} = \frac{\alpha[\nu_1(1+z),\nu_2(1+z),T_{\rm
  e}]}{4\sqrt{\pi}(1+z)^4}
  \frac{\Gamma(3\beta-1/2)}{\Gamma(3\beta)}n_{\rm e0}^2\,r_{\rm c}\,
  \left[1+\left(\frac{\theta}{r_{\rm c}/d_{\rm A}(z)}\right)^2\right]
  ^{-3\beta+\frac{1}{2}},
\end{equation}
and
\begin{equation}
  \label{eqn:betay}
  y(\theta) = \sqrt{\pi}\frac{\Gamma(3\beta/2-1/2)}{\Gamma(3\beta/2)}
  \frac{(k_{\rm B} T_{\rm e}/m_{\rm e} c^2)\sigma_{\rm T}n_{\rm
  e0}r_{\rm c}} {[1+(d_{\rm A}(z)\,\theta/r_{\rm c})^2]^{3\beta/2-1/2}},
\end{equation}
where $r_c$ is the core radius of the cluster (in physical lengths),
$d_{\rm A}(z)$ is the angular diameter distance to the redshift $z$
(e.g., Birkinshaw, Hughes, Arnaud 1991; Kobayashi, Sasaki, Suto 1996).
Therefore we attempt the following fits to the projected profiles of
clusters A and B:
\begin{equation}
  \label{eqn:betafit}
S_{\rm X}(\theta) \propto 
[1+ (\theta/\theta_{\rm c,X})^2]^{-3\beta_{\rm X}+1/2},
\qquad
y(\theta) \propto 
[1+ (\theta/\theta_{\rm c,y})^2]^{-3\beta_{\rm y}/2+1/2}. 
\end{equation}
The results of the fits to equations (\ref{eqn:betamodel}) and
(\ref{eqn:betafit}) are summarized in Figure~\ref{fig:rcbeta}, and
also plotted in Figure \ref{fig:2dprofile} (thick solid and thin
dashed lines for cluster A and B, respectively).  The fits are
performed in the range of $\epsilon_{\rm G} < r < 2r_{\rm vir}$ for
equation (\ref{eqn:betamodel}), and of $\epsilon_{\rm G}/d_{\rm A} <
\theta < 2r_{\rm vir}/d_{\rm A}$ for equation (\ref{eqn:betafit}).
Note that the core radii of both clusters decrease with redshift
indicating that the clusters are still contracting from $z=1$ to $z=0$
and thus are not in equilibrium at $z=1$.

While Figure \ref{fig:2dprofile} clearly illustrates that the separate
fits to the $\beta$-model predictions (eqs.[\ref{eqn:betamodel}] and
[\ref{eqn:betafit}]) are quite successful, the best-fitted values of
$r_c$ (or $\theta_c$) and $\beta$ change substantially depending on
whether one uses $n_e(r)$, $y(\theta)$, or $S_{\rm X}(\theta)$ for the
fit. This should be ascribed to non--isothermality, asphericity and
local clumpiness of clusters.  A closer look at
Figure~\ref{fig:rcbeta} implies that the former is more important for
clusters at $z=0$ (because the result is almost insensitive to the
line-of-sight directions) while asphericity and local clumpiness
dominate for clusters at $z=1$.

To understand the above, we plot the three dimensional profiles both
for $T_{\rm e}(r)$ and $\rho_{\rm gas}(r)$ in Figure
\ref{fig:3dprofile}.  The integrand of $y$ is proportional to $T_{\rm
e}$ (eq.[\ref{eqn:yint}]) while $S_{\rm X}$ is more weakly dependent
on $T_{\rm e}$. Recall that the bolometric emissivity of the thermal
bremsstrahlung is proportional to $\sqrt{T_{\rm e}}$ and the inclusion
of line emission makes the dependence on $T_{\rm e}$ even weaker for
lower temperature clusters. Thus the gas temperature drop in the
central regions, as shown in Figure \ref{fig:3dprofile}, should
increase both $r_c$ and $\beta$ fitted to $y(\theta)$, and to a lesser
extent to $S_{\rm X}(\theta)$ compared with those fitted to $n_e(r)$,
which is consistent with the systematic trends for clusters at
$z=0$. In Figure \ref{fig:3dprofile} different line-of-sight
projections yield a large scatter of $r_c$ and $\beta$ for clusters at
$z=1$ indicating that asphericity and clumpiness are more appreciable.

Figure \ref{fig:xsubcont6} displays the contours of $|\Delta I_{\rm
mm}|$ (eq.[\ref{eqn:dIth}] at 150 GHz), $\Delta I_{\rm submm}$
(eq.[\ref{eqn:dIth}] at 350 GHz), $S_{\rm X}$ (eq.[\ref{eqn:x-ray}] at
2 -- 10 keV band), and the X-ray emission weighted temperature for
clusters A and B. Apparently the contours of $|\Delta I_{\rm mm}|$ and
$\Delta I_{\rm submm}$ are more extended than that of $S_{\rm X}$,
reflecting that the formers are essentially line-of-sight integrals of
$n_e(r)$ rather than of $n_e^2(r)$ for $S_{\rm X}$.  Also they clearly
exhibit that $\Delta I_{\rm mm}$ and $\Delta I_{\rm submm}$ do not
suffer from $(1+z)^{-4}$ diminishing effect, which is quite suitable
for cluster surveys at high redshifts compared with those in optical
and X-rays.

\section{Estimating the Hubble Constant and Peculiar Velocity of Clusters}

\subsection{Hubble Constant}

The determination of the Hubble constant from cluster observation
relies on the prediction of core radius of clusters of galaxies from
the $\beta$--model fit combining the SZ and X--ray fluxes
(eqs.[\ref{eqn:dToverT}], [\ref{eqn:Sx}] and [\ref{eqn:betay}]):
\begin{eqnarray}
  \label{eqn:rcore}
  r_{\rm c,X,est}&=&\frac{[\Delta T(0)/T\,]^2_{\rm obs}}
  {S_{\rm X}(0)_{\rm obs}}
  \frac{\Gamma(3\beta_{\rm fit,X}-1/2)\Gamma(3\beta_{\rm fit,X}/2)^2}
  {\Gamma(3\beta_{\rm fit,X})\Gamma(3\beta_{\rm fit,X}/2-1/2)^2} \nonumber \\
  &\times&\frac{m_{\rm e}^2c^4\alpha[\nu_1(1+z),\nu_2(1+z),T_{\rm e}]}
  {4\pi^{3/2}(1+z)^4\,\sigma_{\rm T}^2\,k_{\rm B}^2\,T_{\rm e}^2}\,t^{-2}(x) ,
\end{eqnarray}
(Silk, White 1978; Birkinshaw, Hughes, Arnaud 1991).

The reliability of the estimated Hubble constant crucially depends on
the relevance of the isothermal $\beta$--model. Following Inagaki,
Suginohara \& Suto (1995), we examine the distribution of the
parameter:
\begin{equation}
  \label{eqn:fh2d}
  f_{H,{\rm 2D}} \equiv 
  \frac{r_{\rm c,X,fit}}{r_{\rm c,X,est}(\beta_{\rm X,fit})}
= \frac{H_{0,{\rm est}}}{H_{0,{\rm true}}}
\end{equation}
for all of the simulated clusters from three orthogonal directions at
$z=0.0$ and $z=1.0$, where $r_{\rm c,X,fit}$ and $\beta_{\rm X,fit}$
are computed by fitting to the profile of $S_{\rm X}(\theta)$
(eq.[\ref{eqn:Sx}]). The second equality in equation (\ref{eqn:fh2d})
holds if the isothermal $\beta$-model is exact and the angular
diameter distance is approximated as $d_{\rm A} \approx cz/H_0$.  For
comparison, we also compute the similar quantity:
\begin{equation}
  \label{eqn:fh3d}
  f_{H,{\rm 3D}} \equiv \frac{r_{\rm c,n_e,fit}}{r_{\rm
  c,n_e,est}(\beta_{\rm n_e,fit})},
\end{equation}
using $r_{\rm c,n_e,fit}$ and $\beta_{\rm n_e,fit}$ fitted to the
three-dimensional (spherically averaged) gas profile
(\ref{eqn:betamodel}). Note that $f_{H,{\rm 2D}}$, rather than
$f_{H,{\rm 3D}}$, should be considered as a measure of the
observational uncertainties of $H_0$ since only $r_{\rm c,X,fit}$ and
$\beta_{\rm X,fit}$ are directly estimated from the X-ray
observations.

Panels (a) to (d) in Figure \ref{fig:allhist} show the histograms of
$f_{H,{\rm 2D}}$ and $f_{H,{\rm 3D}}$ for the nine clusters at
$z=0.05$ and $z=1.0$. It is interesting to note that there is no
significant systematic bias for $f_{H,{\rm 2D}}$ at $z=0$ even though
the values of $r_{\rm c,X,fit}$ and $\beta_{\rm X,fit}$ are fairly
different from $r_{\rm c,n_e,fit}$ and $\beta_{\rm n_e,fit}$. In fact,
the mean values of $f_{H,{\rm 2D}}$ and $f_{H,{\rm 3D}}$ are close to
unity well within the 1$\sigma$ statistical errors.  As first
discussed by Inagaki, Suginohara, \& Suto (1995), any temperature
structure could bias the value of $H_0$ estimated on the basis of
equation (\ref{eqn:rcore}). Although our simulated clusters do show
some temperature structure, they do not seem to be strong enough to
cause significant systematics.  On the other hand, clusters at $z=1.0$
exhibit the systematic underestimate bias of $H_0$ by $\sim 20$\%.
This underestimate would result from the asphericity of the clusters
because the values for the same clusters from three line--of--sight
directions are significantly different.

\subsection{Peculiar Velocity}

The observed SZ flux $\Delta I_{\nu}$ is contributed from both the
thermal and kinematic SZ effects (eqs.[\ref{eqn:dIth}] and
[\ref{eqn:dIkin}]). By using the different spectral dependence of
these effects, we can estimate the peculiar velocity of the cluster
(Sunyaev, Zel'dovich 1980; Rephaeli, Lahav 1991).  Denote the total SZ
flux in two bands, say $\nu=\nu_1$ and $\nu_2$, by
\begin{equation}
  \label{eqn:dItotal}
  \Delta I_j = i_0 [y~g(x_j)-\tau(v_{r}/c)h(x_j)], ~(j=1 ~\mbox{and}~ 2),
\end{equation}
where $x_j=h_p\nu_j/k_{\rm B}T_{\rm CMB}$.  While the $y$-parameter is
observable, $\tau$ is not unless the three-dimensional temperature
profile is known independently. So one may estimate the peculiar
velocity using the X-ray emission weighted temperature along the
line--of--sight:
\begin{equation}
  \label{eqn:xemtel}
  \langle T_X \rangle \equiv 
\frac{\displaystyle 
\int_{-\infty}^{\infty} T_{\rm e}(r) \alpha[T_e] n_{e}^2(r) dl}
{\displaystyle \int_{-\infty}^{\infty} \alpha[T_e] n_{e}^2(r) dl},
\end{equation}
as follows:
\begin{equation}
  \label{eqn:pecvel}
  \frac{v_{r {\rm ,est}}}{c}=\frac{k_{\rm B}\langle T_X \rangle}
  {m_{\rm e}c^2} \frac{g(x_1)\Delta I_2-g(x_2)\Delta I_1}{h(x_1)\Delta
  I_2-h(x_2)\Delta I_1}.
\end{equation}
If clusters are isothermal, then $y=\tau (k_{\rm B}\langle T_X
\rangle) /(m_ec^2)$ and $v_{\rm r,est}$ is identical to $v_r$ in
equation (\ref{eqn:dItotal}).  For real clusters, however, the
temperature profile is not strictly isothermal and estimates from
equation (\ref{eqn:pecvel}) should be different from $v_{r}$ depending
on the degree of non--isothermality. If we introduce a quantify the
degree of non--isothermality of our simulated clusters:
\begin{equation}
  \label{eqn:fv}
  f_v \equiv \frac{\tau}{y}\frac{k_{\rm B}\langle T_X \rangle}{m_ec^2}, 
\end{equation}
then
\begin{equation}
  \label{truevel}
  v_{r,{\rm est}}=v_{r}\,f_v .
\end{equation}
So $f_v$ can be regarded as a correction factor which relates
$v_{r,{\rm est}}$ to the correct peculiar velocity:

We compute $f_v$ for all simulated clusters at $z=0.05$ and $z=1$ from
three orthogonal line--of--sight directions (Figs. \ref{fig:allhist}e
and f). Compared with the histograms of $f_{H,{\rm 2D}}$ and
$f_{H,{\rm 3D}}$, that of $f_v$ is more centrally concentrated. This
is because $f_v$ is sensitive only to the non--isothermal structure
along the line-of-sight while largely free from either the existence
of local clumpiness or asphericity unlike $f_{H,{\rm 2D}}$ and
$f_{H,{\rm 3D}}$. Since $\langle T_X \rangle$ has more weights on the
high-density central regions, $f_v$ should be less than unity for
clusters with central temperature drop as in our simulated
ones. Panels (e) and (f) in Figure \ref{fig:allhist} indicates that
this is the case. In summary, the estimates of $v_r$ are fairly
reliable with overall fractional uncertainties $\sim 10$\% at
$z\approx 0$ and $\sim 20$\% at $z=1$


\section{Conclusions and Discussion}

We have performed a series of numerical simulations with particular
attention to the feasibility of estimating the Hubble constant $H_0$
and the cluster peculiar velocity $v_r$ via multi-band observations.
Let us briefly summarize our conclusions here.

\begin{figure}[tbph]
\begin{center}
   \leavevmode\psfig{file=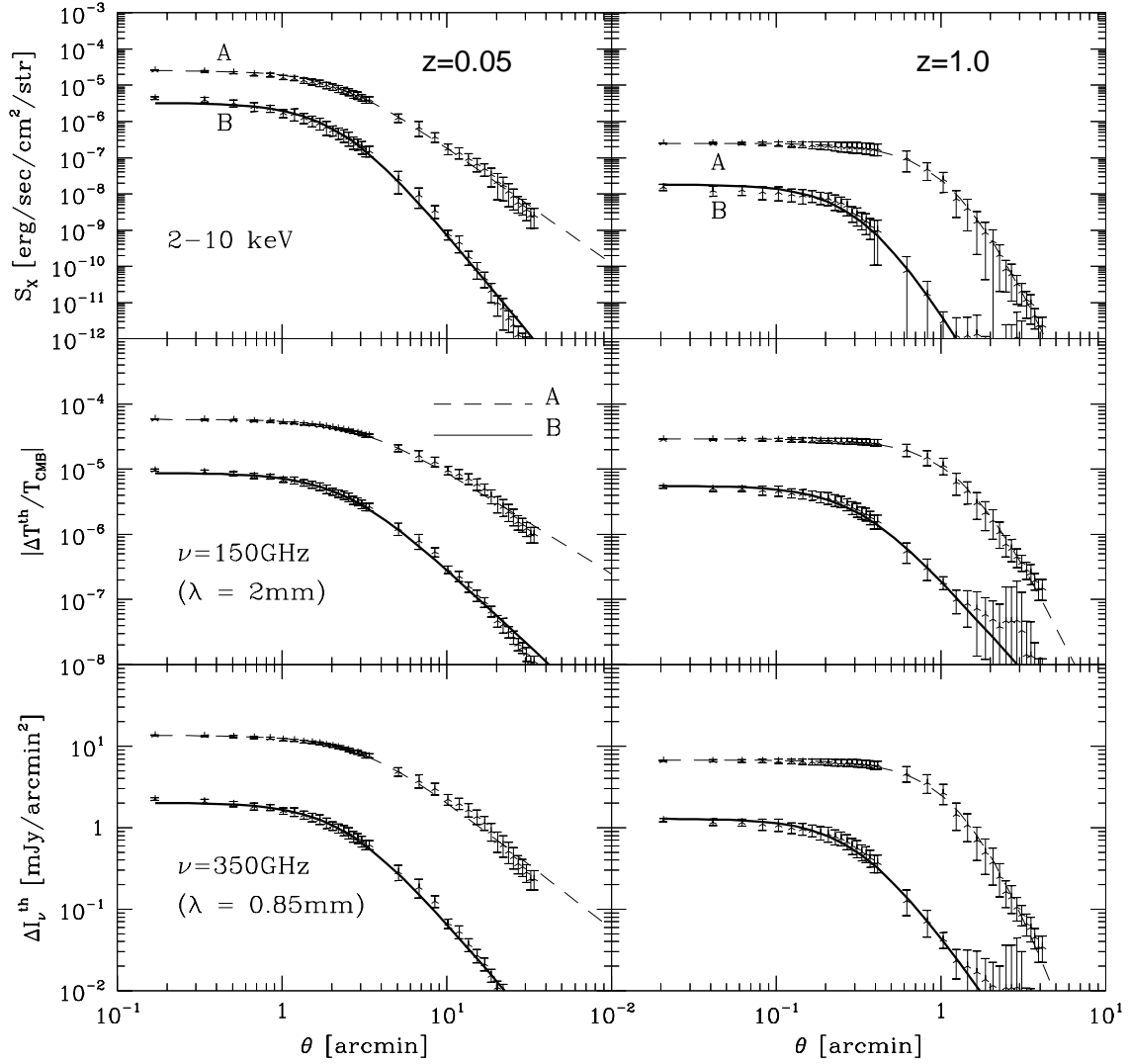,width=15cm}
\end{center}
\caption{Projected profiles of X--ray surface 
brightness ({\it top}), the thermal SZ temperature decrement ({\it
middle}), and the thermal SZ submm surface brightness ({\it bottom})
of clusters A and B at $z=0.05$ ({\it left}) and $z=1.0$ ({\it
right}). The quoted error bars at each angular radius indicate the
1-$\sigma$ statistical dispersion computed from the 16 points on the
projected circle with the radius.
\label{fig:2dprofile}
}
\end{figure}

\begin{figure}[tbph]
\begin{center}
   \leavevmode\psfig{file=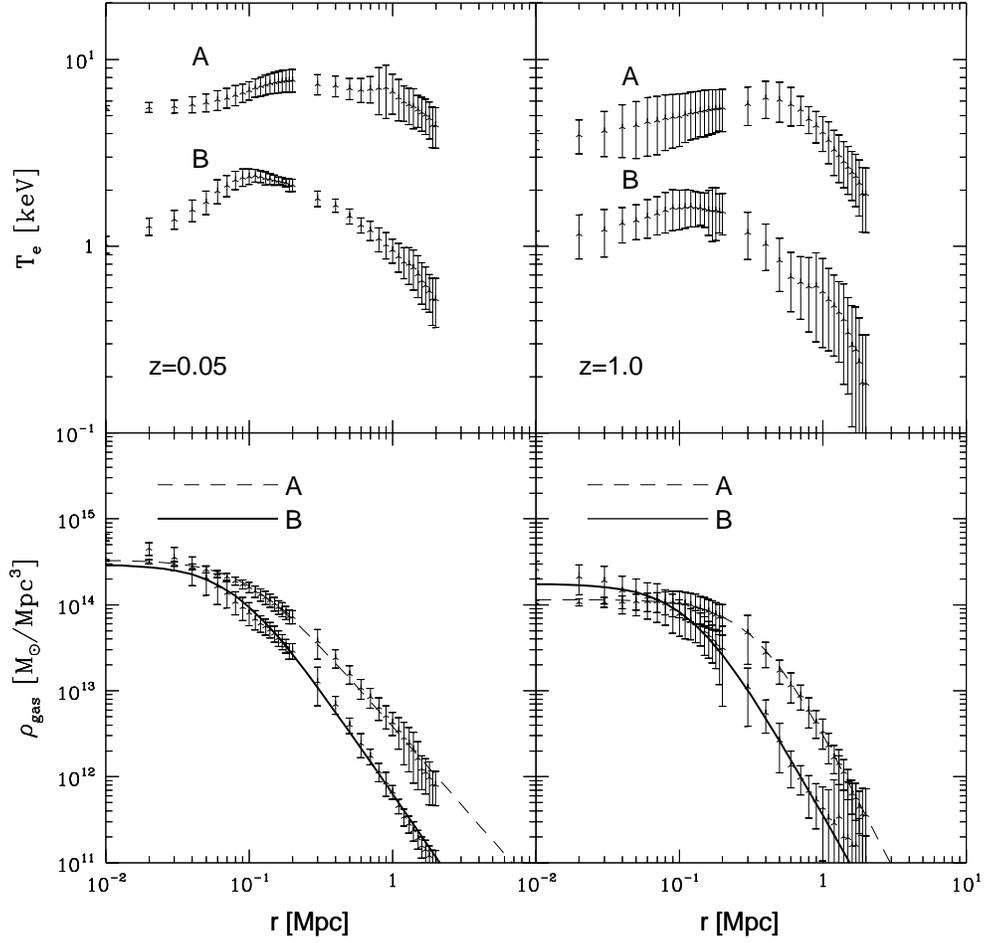,width=15cm}
\end{center}
\caption{Spherically averaged profiles of
gas temperature ({\it upper}) and gas density ({\it lower}) of
clusters A and B at $z=0.05$ ({\it left}) and $z=1.0$ ({\it right}).
The radius $r$ is in physical lengths.
\label{fig:3dprofile}
}
\end{figure}

(i) A conventional isothermal $\beta$-model describes well our
simulated clusters both in projected X-ray surface brightness and the
SZ flux (Fig.\ref{fig:2dprofile}) as well as the three-dimensional gas
density profile (Fig. \ref{fig:3dprofile}).  This is consistent with
the analytical (Makino, Sasaki, Suto 1998) and numerical work (Eke,
Navarro, Frenk 1998) on the basis of the universal density profile of
dark matter halo (Navarro, Frenk, White 1997).

\begin{figure}[tbph]
\begin{center}
   \leavevmode\psfig{file=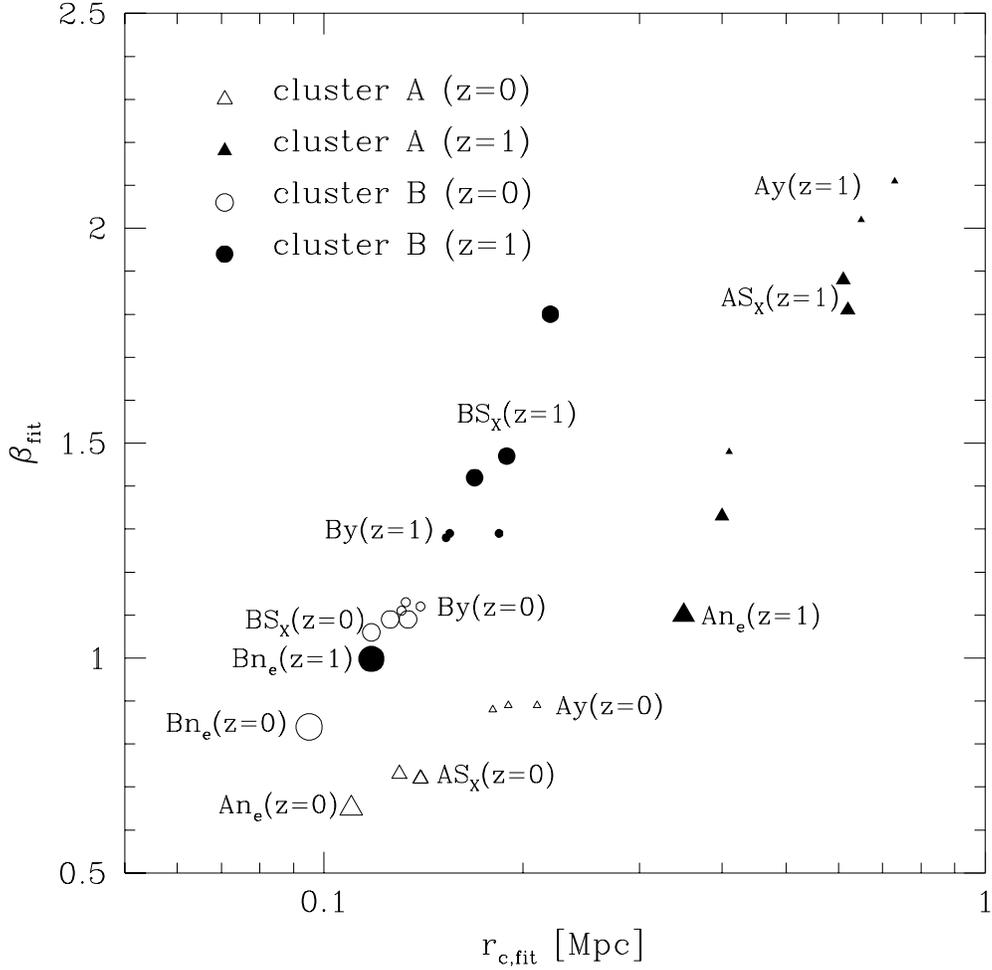,width=15cm}
\end{center}
\caption{ The core radius (in physical lengths) and $\beta$-parameter 
for clusters A (triangles) and B (circles) at $z=0$ (open symbols) and
$z=1$ (filled symbols).  Different sizes of different symbols refer to
the values from separate fits to $n_e(r)$, $S_X$ and $y$-parameter
(from three orthogonal line-of-sight directions for the latter two).
\label{fig:rcbeta}
}
\end{figure}

(ii) The best-fit values for the core radius and $\beta$-parameter
change significantly depending on whether one attempts to fit the
X-ray surface brightness, the SZ flux or the gas density
(Fig.\ref{fig:rcbeta}). This reflects the non-isothermal temperature
profile, local clumpiness and aspherical gas structure.

\begin{figure}
\begin{center}
  \vspace{8cm} 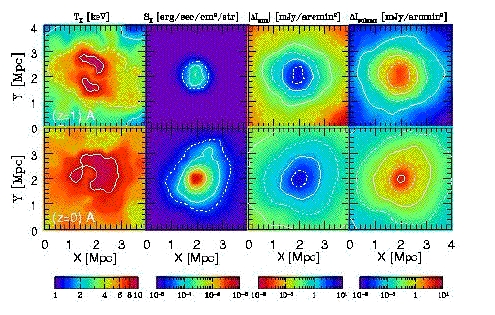
\end{center}
\begin{center}
  \vspace{8cm} 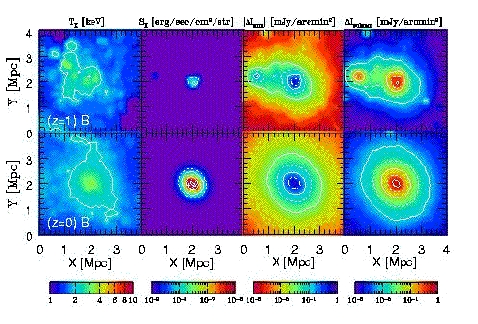
\end{center}
\vspace*{-0.3cm}
\caption{Projected views of clusters A ({\it upper panels}) and 
B ({\it lower panels}) at $z=1$ and $z\approx 0$. A box of (4Mpc)$^3$
(in physical lengths) located at the center of each cluster is
extracted.  The X-ray emission--weighted temperature ($T_X$), X-ray
surface brightness ($S_X$), and the SZ surface brightness at mm and
submm bands ($|\Delta I_{\rm mm}|$ and $\Delta I_{\rm submm}$) are
plotted on the projected X-Y plane by integrating over the
line-of-sight direction (Z). The X and Y coordinates are in the
physical lengths at the corresponding redshift, and related to the
angular coordinate $\theta$ from the cluster center as $\theta
d_A(z)$. At $z\approx 0$, $d_A(z)$ can be replaced by the real
distance to the cluster from the observer.}
\label{fig:xsubcont6}
\end{figure}

\begin{figure}
  \begin{center}
    \vspace{2cm}
    \leavevmode\psfig{file=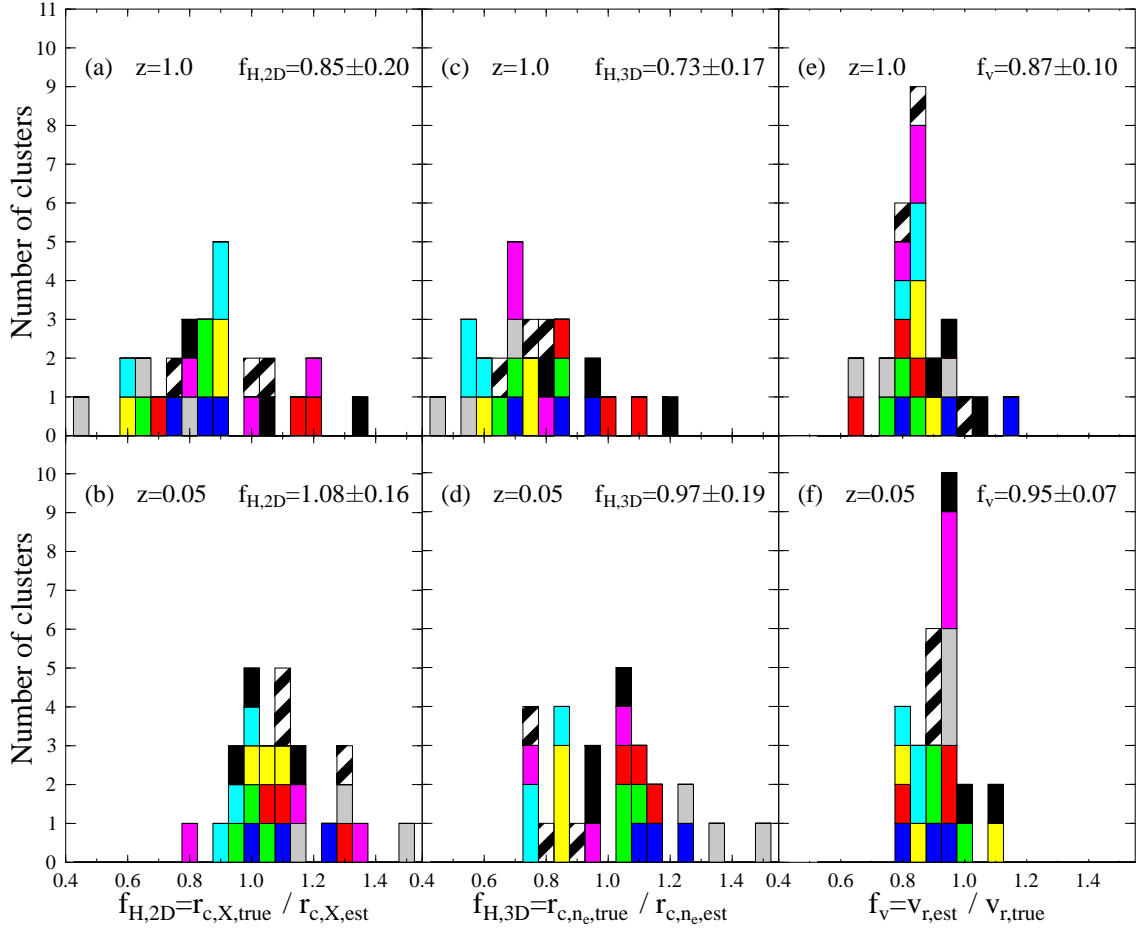,angle=-90,width=15cm}
  \end{center}
\caption{Distribution of $f_{H,{\rm 2D}}$ ({\it left}), 
$f_{H,{\rm 3D}}$ ({\it middle}), and $f_v$ ({\it right}) for all the
simulated clusters (nine in total) at $z=0.05$ and $z=1.0$ viewed from
three different line-of-sight directions. Different patterns of the
histogram correspond to different clusters. The mean and 1$\sigma$
statistical errors are quoted in each panel.
\label{fig:allhist}}
\end{figure}

(iii) Provided that our simulated clusters constitute a representative
sample of the observed clusters, the systematic and statistical errors
in the estimates of $H_0$ and $v_r$ are relatively small
(Fig.\ref{fig:allhist}); fractional uncertainties of the estimates of
$H_0$ amount to $\sim 20$\% mainly due to the departure from the
isothermal and spherical gas density distribution. Those of $v_r$
range $\sim 10$\% at $z\approx 0$ and $\sim 20$\% at $z=1$.  Since the
three-dimensional temperature structure is very difficult to
reconstruct from the projected temperature map (compare
Figs. \ref{fig:3dprofile} and \ref{fig:xsubcont6}), the correction for
the non-isothermality is not easy.

The similar temperature drop in the clusters discussed here was found
earlier in simulations by Evrard (1990) and also reported in some
observed clusters (e.g., Ikebe et al. 1997). Evrard (1990) ascribed
this to the fact that earlier collapsed gas is only mildly shocked
relative to the subsequent infalling gas. Although this is one
reasonable interpretation, it could result simply from the shape of
gravitational potential (Navarro, Frenk, White 1997; Makino, Sasaki,
Suto 1998) or even from the lack of spatial resolution of simulations
to represent the shock in the core regions.  Therefore results from
improved simulations are clearly needed; a larger number of gas and
dark matter particles, $N=128^3$ and even $N=256^3$, is necessary to
resolve the cluster core reliably.  Then it makes sense to incorporate
effects of cooling and heating which have been neglected here mainly
because they are not so important with the mass resolutions of our
present simulations. We have considered only one fairly specific
cosmological model, and it is interesting to examine the cosmological
model dependence of the results.  Nevertheless we hope that our
current results have highlighted several potentially important
implications for the future cluster observations in the X--ray, mm,
and submm bands.

\bigskip

We thank Tetsu Kitayama and Yi-Peng Jing for useful discussions and an
anonymous referee for several constructive comments.  We gratefully
acknowledge the use of two publicly available numerical packages,
Hydra by H.M.P.Couchman, P.A.Thomas and F.R.Pearce, and COSMICS by
E.Bertschinger, for the present simulations. Numerical computations
were carried out on VPP300/16R and VX/4R at the Astronomical Data
Analysis Center of the National Astronomical Observatory, Japan, as
well as at RESCEU (Research Center for the Early Universe, University
of Tokyo) and KEK (National Laboratory for High Energy Physics,
Japan). This research was supported in part by the Grants-in-Aid by
the Ministry of Education, Science, Sports and Culture of Japan
(07CE2002) to RESCEU, and by the Supercomputer Project (No.97-22) of
High Energy Accelerator Research Organization (KEK).

\newpage

\bigskip 

\parskip2pt
\newpage
\centerline{\bf REFERENCES}
\bigskip

\def\apjpap#1;#2;#3;#4; {\pp#1, {#2}, {#3}, #4}
\def\apjbook#1;#2;#3;#4; {\pp#1, {#2} (#3: #4)}
\def\apjppt#1;#2; {\pp#1, #2.}
\def\apjproc#1;#2;#3;#4;#5;#6; {\pp#1, {#2} #3, (#4: #5), #6}

\apjpap Barbosa, D., Bartlett, J.G., Blanchard, A., \& Oukbir,
J. 1996;A\&A;314;13;
\apjppt Bertschinger, E. 1995;astro--ph/9506070;
\apjpap Birkinshaw, M., Hughes, J.P. \& Arnaud, K.A. 1991;
ApJ;379;466;
\apjpap Couchman, H.M.P., Thomas, P.A. \& Pearce, F.R., 1995;
ApJ;452;797;
\apjppt Eke, V.R, Navarro, J.F., \& Frenk, C.S. 1998;
      ApJ, submitted (astro--ph/9708070);
\apjpap Evrard, A.E. 1990;ApJ;363;349;
\apjpap Hattori, M., Ikebe, Y., Asaoka, I., Takeshima, T.,
B{\"o}hringer, H., Mihara, T., Neumann, D.M., Schindler, S., Tsuru, T.
\& Tamura, T., 1997;Nature;388;146;
\apjpap Holzapfel, W.L., Ade, P.A.R., Church, S.E., Mauskopf, P.D.,
Rephaeli, Y., Wilbanks, T.M., \& Lange, A.E. 1997;ApJ;481;35;
\apjpap Ikebe, Y., Makishima, K., Ezawa, H., Fukazawa, Y., Hirayama,
Y., Honda, H., Ishisaki, Y.,Kikuchi, K., Kubo, H., Murakami, T.,
Ohashi, T., Takahashi, T., Yamashita, K. 1997;ApJ;481;660;
\apjpap Inagaki, Y., Suginohara, T.,\& Suto, Y. 1995;PASJ;47;411;
\apjppt Kitayama, T., Sasaki,S., \& Suto, Y. 1998;PASJ, in press
 (astro--ph/9708088);
\apjpap Kitayama, T., \& Suto, Y. 1996;ApJ;469;480;
\apjpap Kitayama, T., \& Suto, Y. 1997;ApJ;490;557;
\apjpap Kobayashi, S., Sasaki,S., \& Suto, Y. 1996;PASJ;48;L107;
\apjpap Makino, N., Sasaki, S., \& Suto, Y. 1998;ApJ;
   497;April 20 issue, in press (astro-ph/9710344);
\apjpap Masai, K. 1984;Ap\&SS;98;367;
\apjpap Navarro, J.F., Frenk, C.S., \& White, S.D.M. 1997;ApJ;490;493;
\apjpap Press, W. H., \& Schechter, P. 1974;ApJ;187;425;
\apjpap Rephaeli, Y. \& Lahav, O. 1991;ApJ;372;21;
\apjpap Rephaeli Y. 1995; ARA\&A; 33; 541;
\apjpap Roettiger, K., Stone, J.M., \& Mushotzky, R.F. 1997;ApJ;482;588;
\apjbook Rybicki, G.B., Lightman, A.P. 1979;Radiative Processes in
Astrophysics; Wiley;New York;
\apjpap Schindler, S., Hattori, M., Neumann, D.M. \& B{\"o}hringer, H., 
 1997;A\& A;317;645;
\apjpap Silk, J. \& White, S.D.M. 1978; ApJ;226;L103;
\apjpap Silverberg, R.F., Cheng, E.S., Cottingham, D.A., Fixsen, D.J., 
Inman, C.A., Kowitt, M.S., Meyer, S.S., Page, L.A., Puchalla, J.L., 
\& Rephaeli, Y. 1997;ApJ;485;22;
\apjpap Sunyaev R.A. \& Zel'dovich Ya.B.  1972;Comments on
Astrophys.\& Space Phys.;4;173;
\apjpap Sunyaev R.A. \& Zel'dovich Ya.B.  1980;MNRAS;190;413;
\apjpap White, S. D. M., Efstathiou, G., \& Frenk, C. S. 1993;
  MNRAS;262;1023;

\clearpage

\clearpage

\clearpage

\clearpage

\clearpage

\end{document}